\documentclass[pre,nofootinbib,superscriptaddress,onecolumn,preprintnumbers]{revtex4}
\usepackage{graphicx}
\usepackage{amsmath,amssymb}
\usepackage{color}

\usepackage{booktabs}

\renewcommand{\eqref}[1]{Eq.(\ref{#1})}

\begin{document}
\title{Kinetic Model for Dark Energy - Dark Matter Interaction:\\
Scenario for the Hubble Tension}

\author{G. Montani}
\email{giovanni.montani@enea.it}
\affiliation{ENEA, Nuclear Department, C. R. Frascati, Via E. Fermi 45,  Frascati, 00044 Roma, Italy}
\affiliation{Physics Department, ``Sapienza'' University of Rome,  P.le Aldo Moro 5, 00185 Roma, Italy}

\author{N. Carlevaro}
\affiliation{ENEA, Nuclear Department, C. R. Frascati, Via E. Fermi 45,  Frascati, 00044 Roma, Italy}

\author{L.A. Escamilla}
\affiliation{School of Mathematics and Statistics, University of Sheffield, Hounsfield Road, Sheffield S3 7RH, United Kingdom}

\author{E. Di Valentino}
\affiliation{School of Mathematics and Statistics, University of Sheffield, Hounsfield Road, Sheffield S3 7RH, United Kingdom}

\begin{abstract}
We analyze a model for Dark Energy - Dark Matter interaction, based on a
decaying process of the former into the latter. 
The dynamical equations are constructed following a kinetic formulation, 
which separates the interacting fluctuations from an equilibrium distribution of both species. The emerging dynamical picture consists of coupled equations, which 
are specialized in the case of a Dark Energy equation of state parameter; we deal with a modified Lambda Cold Dark Matter ($\Lambda$CDM) model, which is investigated versus a possible interpretation of the Hubble tension. Using an optimized set of the model's free parameters, it can be shown that the obtained Hubble parameter can, in principle, address the tension. We then use the most recent datasets from late Universe sources and compressed information from the Cosmic Microwave Background data to constrain the free parameters and compare the addressed scenario to the standard $\Lambda$CDM model. The study outlines how our proposal is preferred by the data in all cases, based on fit quality, while also alleviating the tension.
\end{abstract}

\maketitle

\section{Introduction}
Ever since the original Hubble measurement in 1929, the value of the Hubble constant, $H_0$, has played a central role in experimental and theoretical cosmology. Only in the epoch of the so-called ``precision cosmology'' has it been detected with a good degree of accuracy~\cite{2000Natur.404..955D,2013ApJS..208...20B,Planck:2018vyg}. However, in recent years, the Hubble constant value has been subjected to a surprising fate: clear observational evidence~\cite{Planck:2018vyg,2024ApJ...963L..43A,Scolnic_2022,2018ApJ...859..101S,Brout:2022vxf,DAINOTTI202430,Dainotti2023ApJ...951...63D,Riess:2021jrx,Scolnic:2023mrv,Jones:2022mvo,Anand:2021sum,Freedman:2021ahq,Uddin:2023iob,Huang:2023frr,Li:2024yoe,Pesce:2020xfe,Kourkchi:2020iyz,Schombert:2020pxm,Blakeslee:2021rqi,deJaeger:2022lit,Murakami:2023xuy,Breuval:2024lsv,SPT-3G:2022hvq,ACT:2020gnv,2024A&A...686A.210F,2024arXiv241218493L,2024arXiv241214750S} suggests a possible dependence of $H_0$ on the sources adopted for its determination. This observational discrepancy, known as the ``Hubble tension'', calls attention to an explanation based on a possible unrecognized redshift evolution of astrophysical sources (see, e.g., the analysis in~\cite{2021ApJ...914L..40D,Dainotti2023mnras,2022PASJ...74.1095D,2024PDU....101874}, even if not yet statistically significant) or on possible ``new physics'', as discussed in~\cite{DiValentino:2021izs,2024Univ...10..140C,2023PDU....4001201C,Kamionkowski:2022pkx,Khalife:2023qbu,Abdalla:2022yfr,Perivolaropoulos:2021jda,DiValentino:2020zio,Verde:2023lmm,Giare:2024syw,Giare:2024akf,2024arXiv240814878B,hu-wang03,arXiv:2411.16678,arXiv:2304.01831} and references therein.

In~\cite{Dainotti2021apj-powerlaw} (see also~\cite{Dainottigalaxies10010024,Krishnan:2020obg,Krishnan:2020vaf,kazantzidis,hu-wang02,hu-wang01,2025arXiv250111772D}), it was argued that deviations from the standard Lambda Cold Dark Matter ($\Lambda$CDM) or $w_0w_a$CDM models can also be recovered within the Type Ia Supernova (SN) Pantheon sample (for a critical discussion of the binning approach see \cite{Brout:2020bbg}). Such modifications are well-reproduced by the modification $H_0 \to H_0 f(z)$, with $f(z) \propto (1+z)^{-\alpha}$, where $\alpha$ is a parameter with a value on the order of $10^{-2}$. Such behavior has been properly reproduced via a modified $f(R)$-gravity in the Jordan frame~\cite{Sotiriou-Faraoni:2010,NOJIRI201159,arx1705.11098,arx2307.16308,2011PhR...509..167C}, as investigated in~\cite{schiavone_mnras}. For further studies addressing the Hubble tension via modified gravity theories, see~\cite{Nojiri:2022ski,Odintsov:2020qzd,Petronikolou:2023cwu,FrancoAbellan:2023gec,Ravi:2023nsn}, while for approaches involving evolutionary phantom energy, see~\cite{deangelis-fr-mnras,2024PhRvD.109b3527A}. A recent model utilizing slow-rolling scalar dynamics, which deals with the binned SN data (as processed in~\cite{Dainotti2021apj-powerlaw}), is studied in~\cite{2024PDU....4401486M}. Finally, for a study on the creation of scalar matter via the cosmological gravitational field, see~\cite{erdem24a}.

However, the difficulty in reconciling observational data from Baryon Acoustic Oscillations (BAO)~\cite{eBOSS:2020yzd,DESI:2024mwx} with the SH0ES Collaboration measurements on SN~\cite{Brout:2022vxf} (which live in overlapping redshift regions) calibrated by Cepheids, has served as motivation for an early Universe modification of the dynamics, for instance affecting the sound horizon value, in addition to late Universe models (see~\cite{2021MNRAS.505.3866E,Knox:2019rjx}). In~\cite{2022JHEAp..36...27V,PhysRevD.98.083501,2023Univ....9..393V,2020PhRvD.102b3518V,Allali:2021azp,Anchordoqui:2021gji,Khosravi:2021csn,Clark:2021hlo,Wang:2022jpo,Anchordoqui:2022gmw,Reeves:2022aoi,Yao:2023qve,daCosta:2023mow,Wang:2024dka}, it has been suggested that the most reliable solution to the Hubble tension could require a combination of both late and early Universe modified physics. For studies concerning a possible variation of the matter density parameter, see~\cite{Ziad23,Dainotti2024PDU....4401428D,arx2203.10558,arx2206.11447,arx2304.02718}.

Here, we consider a Dark Energy (DE) - Dark Matter (DM) interaction model (for a review on this topic, see~\cite{2016RPPh...79i6901W}; see~\cite{naidoo2024PhRvD} and~\cite{erdem24a} for an approach to matter creation by a scalar field; see~\cite{Pourtsidou:2016ico,DiValentino:2017iww,Kumar:2017dnp,Yang:2018uae,vonMarttens:2019ixw,Lucca:2020zjb,Zhai:2023yny,Bernui:2023byc,Hoerning:2023hks,Giare:2024ytc,Giare:2024smz,escamilla2023idegp,Benisty:2024lmj,Silva:2024ift,Forconi:2023hsj,DiValentino:2019ffd,arXiv:2409.15513,spin12} for models of DE/DM trying to address the Hubble tension), based on a kinetic approach to describe the decaying process of the DE constituents into DM particles. Thus, we assume that the equilibrium configuration for the species is perturbed by small fluctuations associated with a Boltzmann equation typical of a decay process from one species into another. Actually, while the DE decay process is described via the kinetic theory (by reconstructing the macroscopic equation for the fluctuations), the energy transfer to DM via particle creation is recovered by the conservation of the total energy-momentum tensor for the two species. As a result of this underlying physical scenario, we get a modified Friedmann equation containing an additional evolutionary term with respect to the $\Lambda$CDM model, which is the net effect of the DE decay process. Clearly, when this additional term vanishes, the model exactly reduces to a $\Lambda$CDM picture, corresponding to the instant when the DE-DM interaction starts.

In this paper, we first show how, using an optimized set of the model's free parameters, the Hubble parameter profile can theoretically address the Hubble tension. This occurs because the value of $H_0$ can align with the range of SH0ES measurements, and the curve coincides with the Planck $\Lambda$CDM model~\cite{Planck:2018vyg} for $z > 1$. This theoretical prediction underscores the potential of the proposed physical scenario to reconcile the two conflicting measurements. Additionally, the insensitivity of the Cosmic Microwave Background (CMB) measurements to changes in the angular diameter distance caused by our model should be considered a requirement for resolving the Hubble tension. This suggests that our Hubble parameter for large $z$ can be represented formally as degenerate with Planck $\Lambda$CDM’s predictions, but with different cosmological parameters.

To finalize, we perform a detailed Bayesian parameter inference for both the $\Lambda$CDM model and the proposed one, dubbed DE-DM. We use some of the most recent available datasets for late Universe sources and also a BAO-like implementation of the CMB data. This analysis reveals that, in all the addressed cases, our model is preferred by the data based on fit quality. The Bayes factor indicates that our model is at least as good as (if not slightly better than) the $\Lambda$CDM model, despite the inclusion of three additional free parameters. Our proposal also has the merit of slightly alleviating the Hubble tension in some of the studied cases, although with the addition of the SH0ES prior for the Hubble parameter. Also the scenarios obtained incorporating CMB information are particularly interesting, as the DE-DM model achieves a better data fit and reduces the Hubble tension due to its intrinsic dynamical features.

The paper is structured as follows: in Sec.~2 we introduce the theoretical framework, in Sec.~3 we define the dynamics for the Hubble parameter, in Sec.~4 we describe the methodology used for the data analysis and the results, and finally in Sec.~5 we derive our conclusions.

\section{Theoretical framework}
We consider here a model for a DE-DM interaction, based on the idea that a fraction of the DE constituents decay, resulting in the creation of DM particles (this process is viewed in a kinetic formulation).

In order to describe the decay process of DE particles (labeled by the suffix $DE$), we first consider a flat isotropic Universe, described by the line element
\begin{align}
    ds^2 = dt^2 - a^2(t)\left( 
    dx^2 + dy^2 + dz^2\right) \, , 
    \label{in1}
\end{align}
where $t$ is the synchronous time (we use $c=1$ units), $(x,y,z)$ are Cartesian coordinates, and $a(t)$ is the cosmic scale factor, governing the Universe expansion.
Then, we set the DE distribution function as $f = f_{eq}(E_{DE}) + \delta f(E_{DE})$, where $E_{DE}$ denotes the particle energy, $f_{eq}$ the dominant equilibrium contribution, and $\delta f$ describes the decay process. Thus, while $f_{eq}$ is responsible for the DE density of the Universe, i.e.
\begin{align}
    \rho_{DE} = \rho_{DE}^0 (1 + z)^{3w + 3}\,,
\end{align}
($\rho_{DE}^0$ being its present-day value and $w < -1/3$ the equation of state parameter), $\delta f$ is subjected to the following Boltzmann equation~\cite{bib:montani-primordialcosmology}:
\begin{align}
    \partial_z \delta f + \frac{P_{DE}}{1 + z} \partial_{P_{DE}} \delta f = \frac{\Gamma}{(1 + z)H} \delta f \,,
    \label{in2}
\end{align}
where $P_{DE}$ is the DE particle momentum, $H \equiv a^{-1} \frac{da}{dt}$ is the Hubble parameter, and we are using the redshift time variable $z(t) = \frac{a_0}{a} - 1$ ($a_0$ is the present-day value of the scale factor). We thus get $d(...)/dt = - (1 + z) H \, d(...)/dz$. In the equation above, $\Gamma$ is a constant such that $1/\Gamma$ corresponds to the decay time of the DE species.

We now recall the definition of the energy density fluctuation as:
\begin{align}
    \delta \rho_{DE} \equiv 
    \frac{g_{DE}}{2\pi^2} \int P_{DE}^2 E_{DE} \delta f \; dP_{DE} \, ,
    \label{in3}
\end{align}
where $g_{DE}$ is the particle degree of freedom and we used the isotropy of the momentum space. Hence, we get the following equation:
\begin{align}
    \frac{d \delta \rho_{DE}}{dz} - 
    \frac{3(1 + w)}{1 + z} \delta \rho_{DE} = \frac{\Gamma}{(1 + z)H} \delta \rho_{DE} \, ,
    \label{in4}
\end{align}
where we used the following phenomenological relation for the DE fluctuation pressure $\delta p_{DE}$:
\begin{align}
    \delta p_{DE} \equiv \frac{g_{DE}}{2\pi^2} \int \frac{P_{DE}^4}{3E_{DE}} \delta f \; dP_{DE} = w \, \delta \rho_{DE} \, .
    \label{inx}
\end{align}
This expression, compared with Eq.~(\ref{in3}), for negative values of $w$ (like DE), leads to particles with negative DE squared energy. This does not mean that we are dealing with tachyons but, more realistically, kinetic theory would suggest that the DE constituents are not elementary free particles. Instead, they correspond to relativistic bound states (self-interacting micro-clusters) which can violate the energy condition~\cite{1988PhRvL..61.1446M}. 

The solution of Eq.~(\ref{in4}) reads
\begin{align}
    \delta \rho_{DE} = \delta \rho_{DE}^{0} (1 + z&)^{3w + 3} 
    \exp \left[ \Gamma \int_0^z \frac{dz'}{(1 + z')H(z')} \right] \, ,
    \label{in5}
\end{align}
where $\delta \rho_{DE}^{0} = \delta \rho_{DE}(z=0)$. We require that the 
sum of the DE and DM energy-momentum tensors is conserved, 
which leads to the following equation 
for the DM energy density fluctuation 
$\delta \rho_{DM}$:
\begin{align}
    \frac{d \delta \rho_{DM}}{dz} - \frac{3}{1 + z} \delta \rho_{DM} = 
    - \frac{\Gamma}{(1 + z)H} \delta \rho_{DE} \, . 
    \label{in6}
\end{align}
Now, defining $\delta \rho \equiv \delta \rho_{DE} + \delta \rho_{DM}$, from Eqs.~(\ref{in4}) and (\ref{in6}), we get the following dynamical equation:
\begin{align}
    \frac{d \delta \rho}{dz} - \frac{3}{1 + z} \delta \rho = - \frac{3 |w|}{1 + z} \delta \rho_{DE} = 
     - 3 |w| \delta \rho_{DE}^{0} (1 + z)^{3w + 2}
     \exp \left[ 
    \Gamma \int_0^z \frac{dz'}{(1 + z')H(z')} \right] \, , 
    \label{in7}
\end{align}
where we emphasized the negative values of $w$ for the DE contribution and we made use of the solution in Eq.~(\ref{in5}).

Finally, we observe that the baryonic component of the Universe is not involved in the particle creation process and its energy density $\rho_b$ satisfies the standard equation
\begin{align}
    \frac{d \rho_b}{dz} - \frac{3}{1 + z} \rho_b = 0 \, .
    \label{in8}
\end{align}
In what follows, we denote by $\rho_m = \rho_{DM} + \rho_b$ the total equilibrium value of the matter component of the Universe and we also introduce the radiation density component indicated by $\rho_r$.

\section{Dynamics for the Hubble parameter}
The Friedmann equation corresponding to the scenario depicted above takes the following form:
\begin{align}
    H^2(z) = \frac{\chi}{3} 
    \left( \rho_m+ \rho_r + \rho_{DE} + \delta \rho \right) \, ,
    \label{in9}
\end{align}
where $\chi$ is the Einstein constant (or equivalently the inverse reduced Planck mass when also $\hbar$ is taken equal unity). Let us now introduce the following quantities:
\begin{align}
    \Omega_m^0 \equiv \frac{\chi \rho_m^0}{3 H_0^2} \, , \qquad
    \Omega_r^0 \equiv \frac{\chi \rho_r^0}{3 H_0^2} \, , \qquad
    \Omega_{DE}^0 \equiv \frac{\chi \rho_{DE}^0}{3 H_0^2} \, , \qquad 
    \Delta \equiv \frac{\chi \delta \rho}{3 H_0^2} \, ,
    \label{in10}
\end{align}
where $\rho_m^0$, $\rho_r^0$ and $\rho_{DE}^0$ are the present-day values of the matter, radiation and DE energy densities, respectively, while $H_0$ is the Hubble constant, i.e., $H(z=0)$. If we define accordingly $\Delta_0 = \Delta(z=0)$, we stress that we assume comparable present-day perturbed contributions of DE and DM (i.e., $\delta \rho_{DM}^0 \sim \delta \rho_{DE}^0$). Thus, $\Delta_0$ can be rewritten as
\begin{align}
    \Delta_0 = (1 + \delta) \Delta_{DE}^0 \, , \qquad
    \Delta_{DE}^0 = \frac{\chi \delta \rho_{DE}^0}{3 H_0^2} \, ,
\end{align}
with $\delta$ being a fraction of unity.

In this scheme, Eq.~(\ref{in9}) can be recast as
\begin{align}
H^2(z) = H_0^2 \Big(\Omega_m^0 (1 &+ z)^3 + \Omega_r^0 (1 + z)^4 + 
\Omega_{DE}^0 (1 + z)^{3w + 3} + \Delta(z) \Big) \, , 
\label{in11}
\end{align}
where $\Omega_{DE}^0 = 1 - \Omega_m^0 - \Omega_r^0 - \Delta_0$, to be coupled with Eq.~(\ref{in7}), here restated in the form
\begin{align}
    \frac{d \Delta}{dz} = \frac{3}{1 + z} \Delta &- 
    3 |w| \Delta_{DE}^0 (1 + z)^{3w + 2} e^{F(z)} \, ,
    \label{in12}
\end{align}
where we have defined $\bar{\Gamma} \equiv \Gamma / H_0$ and we have introduced the auxiliary function $F(z)$ defined as
\begin{align}
    F(z) =  \int_0^z \frac{\bar{\Gamma} (1 + z')^{-1} dz'}{ H(z') / H_0}\,.    \label{in12Fz}
\end{align}
Eqs.~(\ref{in11}), (\ref{in12}) and (\ref{in12Fz}) formally describe the evolution of the Hubble parameter.

\subsection{Modified $\Lambda$CDM model}
In order to develop a dynamical scheme close to the $\Lambda$CDM model, we study the case where $w = -1$, i.e., we set $\rho_{DE} = \text{const}$. It is immediate to specialize Eqs.~(\ref{in11}), (\ref{in12}) and (\ref{in12Fz}) to this scenario. The system can be rewritten as:
\begin{align}
H(z) = H_0 \; \Big(&\Omega_m^0 (1 + z)^3 +\Omega_r^0 (1 + z)^4+
(1 - \Omega_m^0-\Omega_r^0 - \Delta_0) + \Delta(z) \Big)^{1/2} \, , 
    \label{in20}
\end{align}
with
\begin{align}
&\frac{d \Delta}{dz} = 3 (1 + z)^{-1} \big( \Delta(z) - \Delta_{DE}^0 \; e^{F(z)} \big) \, , \label{in21}\\
&\frac{d F}{dz} = \bar{\Gamma}(1 + z)^{-1} \Big(\Omega_m^0 (1 + z)^3 +\Omega_r^0 (1 + z)^4+
(1-\Omega_m^0-\Omega_r^0-\Delta_0) + \Delta(z) \Big)^{-1/2} \, , \label{in22}
\end{align}
where, from Eq.~(\ref{in12Fz}), $F(z=0) = 0$ and we recall that $\Delta_0 = (1 + \delta) \Delta_{DE}^0$. In this scenario, the free parameters of the DE-DM model are $\{ H_0,\, \Omega_m^0,\, \Omega_r^0,\, \Delta_{DE}^0,\, \bar{\Gamma},\, \delta \}$.

From Eq.~(\ref{in21}), $\Delta$ can assume a decreasing behavior in $z$ for a sufficiently large $z$. If we indicate by $z_{\text{in}}$ the instant when $\Delta$ vanishes, then it will increase up to the present-day value $\Delta_0$. We regard $z_{\text{in}}$ as the instant in the past when the decaying process of DE into DM starts. Consequently, at $z \geq z_{\text{in}}$, the process is not present, i.e., $\Delta(z \geq z_{\text{in}}) = 0$, and the model reduces to $\Lambda$CDM-like dynamics of the Universe.\footnote{It is also important to remark how the system above can be rewritten in a slightly different (but interesting) form. Introducing $\tilde{\Omega}(z) = \Omega_m^0 + \frac{\Delta(z)}{(1 + z)^3}$ (and coherently $\tilde{\Omega}_0 = \Omega_m^0 + \Delta_0$), we easily get
\begin{align}
H(z) = H_0 \; \big(\tilde{\Omega}(z)(1 + z)^3 + \Omega_r^0 (1 + z)^4+ 1\!-\! \tilde{\Omega}_0\!-\!\Omega_r^0 \big)^{1/2}\,,
\label{in20bis}
\end{align}
with
\begin{align}
&\frac{d\tilde{\Omega}}{dz} = -3(1 + z)^{-4}\;(1 + \delta)^{-1}(\tilde{\Omega}_0 - \Omega_m^0) e^{F(z)}\,,\label{in21bis}\\
&\frac{dF}{dz} = \frac{(1 + z)^{-1} \bar{\Gamma}}{\big(\tilde{\Omega}(z)(1 + z)^3 + \Omega_r^0 (1 + z)^4+ 1\!-\! \tilde{\Omega}_0\!-\!\Omega_r^0 \big)^{1/2}}\;.\label{in22bis}
\end{align}
In other words, our model can be restated as a $\Lambda$CDM paradigm with an effective matter contribution $\Omega_m(z)$.} We underline that a more detailed analysis of the physical processes triggered at $z_{\text{in}}$ would require a deeper understanding of the microscopic nature characterizing the DE quantum constituents (which is beyond the scope of this paper). However, some important points can be stated to provide more physical insights. 
First of all, we note that the macroscopic equations (\ref{in4}) and (\ref{in6}) for the dynamics of the DE and DM energy density fluctuations, respectively, can be derived without a detailed kinetic formulation by simply postulating a rate of energy transfer from the first to the second species and then requiring the conservation of their energy-momentum tensor.
Thus, this starting point of our analysis is reliably well-posed and makes the dynamical implications rather robust. Clearly, our formulation does not specify the details of the DE-DM dynamics, without providing a Lagrangian density for their interaction. This perspective of a fundamental theory is, in principle, viable, as discussed in \cite{2016RPPh...79i6901W}. However, for both the dark Universe components, no assessed fundamental field description exists and, thus, any specific \textit{ansatz} would bring into the dynamics its own particular features and, this way, the model generality would be limited. Here, we prefer to consider only a basic statistical point of view, resulting into a rate of energy transfer from DE to DM. This scenario is very general and should, in this sense, be able to capture a corresponding rather general dynamical impact. Furthermore, we stress that that the instant $z_{\text{in}}$ depends on the value of the decaying DE rate $\Gamma$. When the Universe is at a sufficiently small redshift (late times), the decaying process proceeds as a transition phase, i.e., a significant amount of DE is transferred to DM as particles very abruptly. The marker of this transition phase is recognized in the non-differentiable nature of the sum of the two perturbed energy densities at the instant $z_{\text{in}}$. After that triggering, the subsequent evolution is smooth (see below).

We now discuss in some detail the physical meaning of the parameter $\Gamma$. For the decaying process to occur at equilibrium (\emph{de facto} being efficient), we require $\bar{\Gamma} \gtrsim 1$. In this respect, we observe that since DE approximately starts to dominate for the last 5 billions of years \cite{melchiorriDE,2008ARA&A..46..385F}, it is reasonable to expect that the decaying time of DE should be a fraction of the Universe's age, greater than $0.3$, to avoid the complete decay of the DE contribution before $z=0$. However, due to the fact that our deviation from the $\Lambda$CDM model concerns the very late Universe dynamics, we will let it vary from 1 to 5. We also observe that the quantity $\Delta_0$ is related to the critical parameter for the present-day DE fluctuations. According to the perturbation scheme adopted above, we must ensure that this contribution, i.e., $\delta\rho_{DE}^0$, is sufficiently small in comparison to the DE equilibrium component $\rho_{DE}^0$ of the present Universe. 

We conclude this subsection by underlining, for completeness in view of the comparisons in the next sections, that a base flat $\Lambda$CDM model profile for $H(z)$ can be simply written from Eq.~(\ref{in20}) with $\Delta = 0$. Moreover, we also introduce the following notation for the flat $\Lambda$CDM model parameters constrained by the CMB data from Planck~\cite{Planck:2018vyg}: $H_{0\textrm{P}}= 67.4 \pm 0.5$, $\Omega_{0\textrm{P}}^m= 0.315 \pm 0.007$, $\Omega_{0\textrm{P}}^r = 9.16 \times 10^{-5} \pm 5 \times 10^{-7}$; and by the Type Ia SN data calibrated with Cepheids from the SH0ES collaboration~\cite{Riess:2021jrx}: $H_{0\textrm{S}} = 73.04 \pm 1.04$, $\Omega_{0\textrm{S}}^m = 0.331 \pm 0.019$, while $\Omega_{0\textrm{S}}^r$ is negligible in this low redshift range (here and in the following, $H_0$ is in units of km s$^{-1}$ Mpc$^{-1}$).

\subsection{Theoretical discussion}
\begin{figure}[ht!]
\centering
\includegraphics[width=9.5cm]{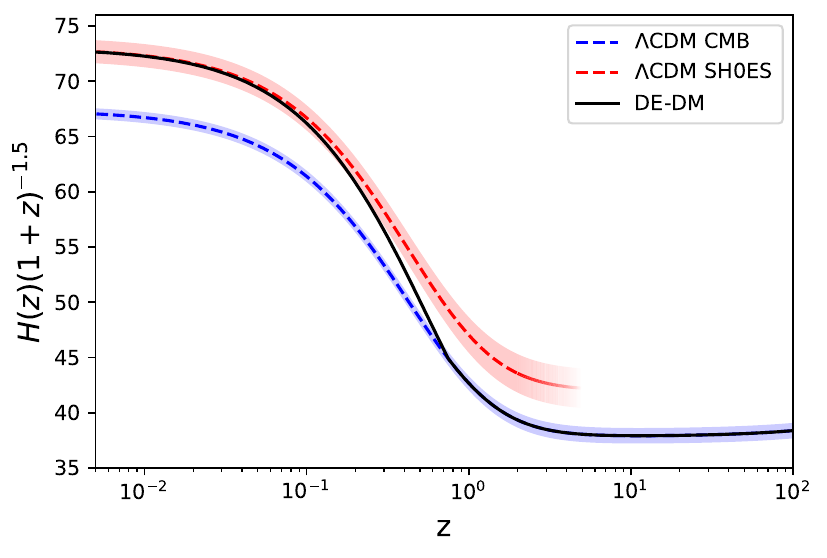}
\caption{Plot of the (normalized) Hubble parameter $H(z)$ from Eqs.~(\ref{in20}), (\ref{paropt1}) and (\ref{paropt2}) (black), and the two $\Lambda$CDM models with parameters constrained by CMB (blue) and, locally, by SH0ES (red), with the corresponding errors. (For interpretation of the references to color in this figure legend, the reader is referred to the web version of this article.)}
\label{figHzth}
\end{figure}

In order to illustrate the potential capability of the proposed physical scenario to connect the incompatible measurements of $H_0$, in Fig. \ref{figHzth} we plot until $z=100$ the Hubble parameter in Eq.~(\ref{in20}) specified for an optimized set of free parameters:
\begin{align}
    &H_0=H_{0\textrm{S}}\,,\quad
   \Omega_m^0=0.268\,,\quad \Omega_r^0=7.834\times10^{-5}\,,\label{paropt1}\\
    &\Delta_{DE}^0=0.124\,,\quad
    \bar{\Gamma} = 1.9\,,\quad
    \delta = 0.2\,,\label{paropt2}
\end{align}
which yield $\Delta_0=0.148$.
It is immediate to see how this profile qualitatively solves the Hubble tension since the value of $H_0$ is in the range of the SH0ES measurement and the curve overlaps the Planck $\Lambda$CDM model for $z > z_{\text{in}} \simeq 1$. In fact, in the figure, we also plot (for reference) the two standard $\Lambda$CDM models constrained by CMB data from Planck (dubbed $H_{\textrm{$\Lambda$CDM}}^{\textrm{CMB}}$) and, locally, by SN calibrated with Cepheids from the SH0ES Collaboration. In this case, the parameters are set imposing $H(z>z_{\text{in}})\simeq H_{\textrm{$\Lambda$CDM}}^{\textrm{CMB}}(z>z_{\text{in}})$, thus when $\Delta(z)=0$. By simple algebra, we obtain $\Omega_m^0\simeq\Omega_{0\textrm{P}}^m H_{0\textrm{P}}^2/H_0^2$, $\Omega_r^0\simeq\Omega_{0\textrm{P}}^r H_{0\textrm{P}}^2/H_0^2$ and finally $\Delta_0\simeq1-\Omega_m^0-\Omega_r^0-(1-\Omega_{0\textrm{P}}^m-\Omega_{0\textrm{P}}^r)H_{0\textrm{P}}^2/H_0^2$, which corresponds to the values in Eqs.~(\ref{paropt1}) and (\ref{paropt2}). The other free parameters are suitably taken following the phenomenological details previously described. We stress that the relations above have only a qualitative significance, since the parameter determination should come from a suitable statistical data analysis. Anyway, it is remarkable that increasing the Hubble parameter in the late Universe implies a corresponding decrease in the critical matter parameter value. This result is actually independent of the model details while relying on the vanishing behavior of $\Delta(z)$ from a given instant.

If the CMB data were to be analyzed using the form of $H(z)$ in Eq.~(\ref{in20}), the parameter inference should, in principle, provide values close to those assumed in this theoretical modeling, since, in this way, $H(z)$ and $H_{\textrm{$\Lambda$CDM}}^{\textrm{CMB}}(z)$ are nearly degenerate for $z>z_{\text{in}}$. However, this analysis is outside the scope of the present paper. The viability of this scenario must thus be guaranteed by the undetectable character of the additional late (say, for $z<z_{\text{in}}$) Universe contribution by the Planck satellite. The most precisely measured parameter by Planck \cite{Planck:2018vyg} turns out to be the acoustic angular scale $\theta_* = r_*/D_M^* = 0.0104109 \pm 0.0000030$, where $r_* = 144.57 \pm 0.22$ Mpc denotes the comoving sound horizon at the time of last scattering $z_* = 1089.80 \pm 0.21$ (recombination), while $D_M^*$ is the comoving angular diameter distance evaluated at the same instant:
\begin{align}
D_M^*=\int_0^{z^*}\frac{c\,dz}{H(z)}\;.
\end{align}
Late Universe models should thus maintain unchanged (within the errors) the value of $\theta_*$, also not varying the sound horizon $r_*$, which is evaluated as an integral from $z_*$ to $\infty$. We stress that a direct quantitative comparison would require a parameter inference analysis directly on the CMB data, thus falling outside the scope of this work. We can evaluate the expression $D_M^*$ using Eq.~(\ref{in20}), which must fall in the range $D_{M\textrm{P}}^* \pm \Delta D_{M\textrm{P}}^*$, where $D_{M\textrm{P}}^*$ is calculated using the CMB-constrained parameters introduced above for a base flat $\Lambda$CDM model. Despite the degree of approximation, this comparison provides good information about the model consistency, i.e., ensuring minimal modifications of early Universe physics. If we evaluate the error $\Delta D_{M\textrm{P}}^*$ using the standard error propagation variance formula, the form of $H(z)$ for the optimized parameters gives $D_M^* = 13735\,\text{Mpc}$, which is compatible with the Planck value $D_{M\textrm{P}}^* = 13864 \pm 168\,\text{Mpc}$.

\section{Data analysis: comparison against $\Lambda$CDM model}
Let us now address a parameter inference procedure performed on our model with the main cosmological data at the background level. The statistical analysis of our model's parameter space was conducted using the (modified) publicly available Bayesian inference code \texttt{SimpleMC}~\cite{simplemc}, which has \texttt{dynesty}~\cite{speagle2020dynesty} implemented (a nested sampling Python library used to compute the Bayesian evidence). The use of this sampling algorithm allows us to compare our model against $\Lambda$CDM and see if the extra degrees of freedom prove to be a hindrance. To this end, we will make use of Jeffrey's scale~\cite{Trotta:2008qt}, which indicates if the evidence against our model is: inconclusive if $0 < \ln B_{\Lambda \text{CDM},i} < 1.0$; weak if $1.0 < \ln B_{\Lambda \text{CDM},i} < 2.5$; moderate if $2.5 < \ln B_{\Lambda \text{CDM},i} < 5.0$; and strong if $5.0 < \ln B_{\Lambda \text{CDM},i}$. If $\ln B_{\Lambda \text{CDM},i} < 0.0$, then we would have evidence in favor of our model (which can also be inconclusive, weak, moderate, and strong) despite presenting a greater number of free parameters than $\Lambda$CDM.

To perform the Bayesian parameter inference, the following datasets were used:
\begin{itemize}
\item[-] The Pantheon+ data release~\cite{Scolnic_2022}, which consists of 1701 light curves of 1550 Type Ia supernovae. When used, it will be written as ``SN'' in the datasets and it spans a redshift range of $0.01 < z < 2.26$.    
\item[-] Measurements of BAO, which include the SDSS Ga-laxy Consensus, quasars, and Lyman-$\alpha$ forests~\cite{eBOSS:2020yzd}. The sound horizon is calibrated using BBN~\cite{Cooke:2013cba}. These datasets are comprehensively detailed in Table 3 of~\cite{eBOSS:2020yzd}. In this work, we will collectively refer to this set of measurements as ``SDSS''.
\item[-] Measurements of BAO distances from the first year of the Dark Energy Spectroscopic Instrument (DESI) \cite{DESI:2024mwx}. We will refer to this dataset as ``DESI''. This dataset will not be used in tandem with SDSS to avoid the risk of double-counting.
\item[-] A likelihood which contains information about the Cosmic Microwave Background (CMB) from Planck, but treated as a ``BAO data point'' at a redshift of $z \sim 1100$. As detailed in~\cite{BOSS:2014hhw}, the background-level information of this ``CMB BAO'' can be encapsulated by the three parameters: $w_b$ (physical baryon density parameter), $w_m$ (physical matter density parameter), and $D_A(\sim 1100)/r_d$. This dataset will be referred to as ``Planck''.
\end{itemize}

Along with these datasets we will also make use of a Gaussian prior for the Hubble parameter $H_0 = 73.04 \pm 1.04$ from SH0ES~\cite{Riess:2021jrx}. Not all cases will include this prior, so when it is used, we will write ``SH0ES'' along with the datasets used. 
For the flat priors of the standard-model's parameters, we have: $\Omega_{m}^{0} = [0.1, 0.5]$, $\Omega_{\rm b} h^2 = [0.02, 0.025]$ for the baryon density, and $h = [0.4, 0.9]$ for the dimensionless Hubble constant, defined as
\begin{align}
h = \frac{H_0}{100\; {\rm km\,s}^{-1}\,{\rm Mpc}^{-1}}\;.
\end{align}
Along with the aforementioned parameters we will also include the derived parameter $\Omega_{r}^{0}$ which corresponds to the radiation's energy density contribution to the Friedmann equation (we refer to it as a ``derived parameter'' since its value can be inferred through the CMB temperature, the effective number of neutrino species and $h$). An important note with regards to $\Omega_{r}^{0}$ is that we will only consider it when using the ``Planck'' likelihood, since its contribution is completely negligible at the low-redshift regime (due to the factor of $a^{-4}$ and its expected present-day value of $8\times10^{-5}$). For our phenomenological specific parameters, we use: $\Delta_{DE}^{0} = [0.05, 0.65]$, $\bar{\Gamma} = [1.0, 5.0]$, and $\delta = [0.1, 1.5]$.

\subsection{Results}
In this subsection, we discuss the results of the parameter inference procedure. In Table \ref{table_params}, we present the datasets used, the inferred parameter values, the Bayes' factor ($\ln B_{\Lambda \mathrm{CDM}, i}$), and $-2\Delta\ln \mathcal{L}_{\rm max}$, which informs us about how well the model fits the data.

Surprisingly, despite the extra degrees of freedom, the Bayes' factor shows, at worst, weak evidence against our model. In some cases, it even has a negative value, indicating inconclusive evidence in favor of our model.

\begin{table*}[th!]
\centering
\footnotesize
\scalebox{0.85}{
\begin{tabular}{lccccccccc}
\hline
Model &  Dataset & $h$ & $\Omega^{0}_{m}$ &$\Omega^{0}_{r}$ & $\delta$ & $\Delta_{DE}^{0}$ & $\bar{\Gamma}$ & $\ln B_{\Lambda \text{CDM},i}$ & $-2\Delta\ln \mathcal{L_{\rm max}}$ \\
\hline
$\Lambda$CDM & SN+SDSS & 0.704 (0.022) & 0.313 (0.014) & - & - & - & - & - & - \\
             & SN+DESI & 0.682 (0.023) & 0.311 (0.012) & - & - & - & - & - & - \\
             & SN+SDSS+SH0ES & 0.719 (0.016) & 0.313 (0.013)  & - & - - & - & - & - & - \\
             & SN+DESI+SH0ES & 0.708 (0.017) & 0.314 (0.013)  & - & - & - & - & - & - \\
\hline
DE-DM   & SN+SDSS & 0.647 (0.052) & 0.277 (0.039)  & - & 0.78 (0.35) & 0.139 (0.075) & 1.73 (0.55) & 0.15 (0.41)  & -3.27 \\
             & SN+DESI & 0.665 (0.054) & 0.299 (0.037)& - & 0.73 (0.32) & 0.125 (0.069) & 2.22 (0.82) & 0.52 (0.41) & -4.79 \\
             & SN+SDSS+SH0ES & 0.717 (0.018) & 0.314 (0.016)& - & 0.534 (0.27) & 0.129 (0.074) & 2.11 (0.75) & 2.14 (0.42) & -1.89 \\
             & SN+DESI+SH0ES & 0.721 (0.017) & 0.322 (0.013)& - & 0.52 (0.31) & 0.17 (0.11) & 2.32 (0.93) & -0.21 (0.39) & -6.35 \\
\hline
$\Lambda$CDM & Planck+SN+SDSS & 0.677 (0.005) & 0.311 (0.007)& $\big(8.39 (0.04) \big)\times 10^{-5}$ & - & - & - & - & - \\
             & Planck+SN+DESI & 0.674 (0.005) & 0.313 (0.007)& $\big(8.43^{+0.07}_{-0.09} \big)\times 10^{-5}$ & - & - & - & - & - \\
             & Planck+SN+SDSS+SH0ES & 0.678 (0.005) & 0.309 (0.007)& $\big(8.41 (0.02) \big)\times 10^{-5}$ & - & - & - & - & - \\
             & Planck+SN+DESI+SH0ES & 0.678 (0.005) & 0.309 (0.007)& $\big(8.34^{+0.13}_{-0.06} \big)\times 10^{-5}$ & - & - & - & - & - \\
\hline
DE-DM   & Planck+SN+SDSS & 0.685 (0.007) & 0.304 (0.008)& $\big(8.28^{+0.09}_{-0.13}\big)\times 10^{-5}$ & 0.69 (0.31) & 0.114 (0.04) & 2.22 (0.78) & -0.18 (0.44) & -5.04 \\
             & Planck+SN+DESI & 0.684 (0.007) & 0.306 (0.007)& $\big(8.31 (0.01)\big)\times 10^{-5}$ & 0.82 (0.35) & 0.094 (0.034) & 2.72 (0.89) & 0.51 (0.43) & -5.12 \\
             & Planck+SN+SDSS+SH0ES & 0.689 (0.007) & 0.301 (0.007)& $\big(8.22 (0.01)\big)\times 10^{-5}$ & 0.69 (0.32) & 0.112 (0.051) & 2.33 (0.75) & -0.41 (0.44) & -6.45 \\
             & Planck+SN+DESI+SH0ES & 0.689 (0.006) & 0.301 (0.007)& $\big(8.18^{+0.13}_{-0.06}\big)\times 10^{-5}$ & 0.79 (0.35) & 0.099 (0.038) & 2.74 (0.89) & -0.14 (0.46) & -6.42 \\
\hline
\end{tabular}
}
\caption{In this table, we report the dataset combinations used, the obtained mean values of some parameters with their $1\sigma$ uncertainties, the Bayes' factor $\ln B_{\Lambda \text{CDM},i}$, and the difference in fitness $-2\Delta\ln \mathcal{L_{\rm max}}$. These last two columns provide a way to quantify how much better or worse a model is at explaining and fitting the data compared with the fiducial model (in this case, the comparison is against $\Lambda$CDM).\label{table_params}}
\end{table*}

When examining the $-2\Delta\ln \mathcal{L}_{\rm max}$ values, we find that our model consistently provides a better fit to the data compared to $\Lambda$CDM, regardless of the dataset used (the more negative the value, the better the fit). This is expected given the three additional parameters in our model. Two cases are particularly noteworthy: Planck + SN + SDSS + SH0ES and Planck + SN + DESI + SH0ES. These combinations show the largest differences in $-2\Delta\ln \mathcal{L}_{\rm max}$, which we attribute to our model's distinctive feature of an increase in $H(z)$ at late times, coupled with the use of the Planck likelihood and the SH0ES prior. The Planck likelihood constrains the physical baryon and matter density parameters $w_b$ and $w_m$, which is reflected in $\Omega_{m}^{0}$ (as seen in its posterior in Table \ref{table_params}). This, in turn, imposes a constraint on the parameter $h$, which for $\Lambda$CDM means an almost fixed value of $h \approx 0.677$ even with the SH0ES prior. This constraint does not apply to our model, as seen by the values of $h$ in Table \ref{table_params}, which have sufficient freedom at late times to allow higher values. This flexibility translates to a better fit for the SN and BAO data and a better Bayes' factor. Moreover, by analyzing the evolution of the function $\Delta(z)$ for all the cases of interest, we find $z_{\text{in}} \simeq 1$. We recall that this corresponds to the instant at which $\Delta = 0$. It is important to note that the best-fit values that will be discussed below do not necessarily correspond to the mean values reported in Table \ref{table_params}.

\subsubsection{SN+DESI+SH0ES}
\begin{figure}[ht!]
\centering
\includegraphics[width=9.5cm]{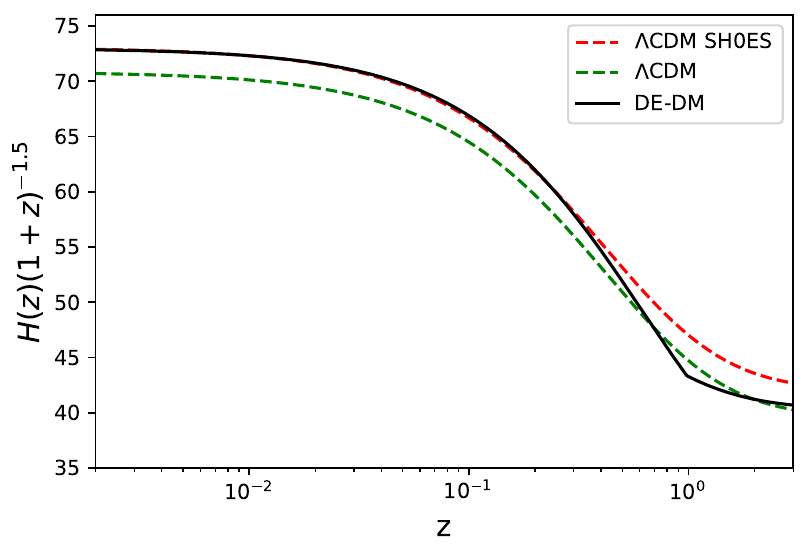}
\caption{SN+DESI+SH0ES --- Evolution of the (normalized) Hubble parameter $H(z)$ from Eqs.~(\ref{in20}), (\ref{parfit1}) and (\ref{parfit2}) (black). The green dashed curve corresponds to the best fit for the base $\Lambda$CDM model, while the red dashed profile denotes the $\Lambda$CDM model constrained by SH0ES. (For interpretation of the references to color in this figure legend, the reader is referred to the web version of this article.)}
\label{figHzfit}
\end{figure}

For this case, we obtain the following best-fit values:
\begin{align}
&H_0 = 72.99\,,\quad
\Omega_m^0 = 0.31\,,
\label{parfit1}\\
&\Delta_{DE}^0 = 0.25\,,\quad
\bar{\Gamma} = 1.70\,,\quad
\delta = 0.28\,,\label{parfit2}
\end{align}
which provide a $\Delta_0 = 0.32$ (we have coherently neglected the radiation contribution in this local analysis). At the same time, the best fit for a base $\Lambda$CDM model provides: $H_0 = 70.85$ and $\Omega_m^0 = 0.31$. The obtained profile of the Hubble parameter is reported in Fig.~\ref{figHzfit}, together with the $\Lambda$CDM counterpart. We also overplot the profile of the $\Lambda$CDM model constrained by the SN dataset from SH0ES, i.e., with $H_{0\textrm{S}}$ and $\Omega_{0\textrm{S}}^m$ as introduced before.

It is immediate to recognize that the local late Universe dynamics for $z<1$ provides a very good agreement with the parameters measured by SH0ES. Moreover, for $z \simeq 1$, our Hubble parameter touches the corresponding $\Lambda$CDM curve, which is an excellent marker of the capability of the underlying physics to alleviate or solve the Hubble tension, even though we are using a SH0ES prior in the data analysis procedure. The obtained profile of the Hubble parameter is, in this respect, more convincing as a solution to the tension than other proposals in the literature, particularly among background-only proposals~\cite{DiValentino:2021izs,Kamionkowski23}.

This analysis in the local late Universe has the significant merit of demonstrating a possible reconciliation of the Cepheid-calibrated SN value of $H_0$ and Planck's CMB one. In fact, it is well-known how these two classes of sources seem to correspond to the $\Lambda$CDM best fit with incompatible values of $H_0$. In particular, BAO data appear well compatible with the Planck satellite $\Lambda$CDM best-fit data. It is precisely the presence of the function $\Delta$ that allows the required modulation of the Hubble parameter, making it flexible enough to interpolate apparently incompatible data.

\subsubsection{Planck+SN+DESI+SH0ES}
\begin{figure}[ht!]
\centering
\includegraphics[width=9.5cm]{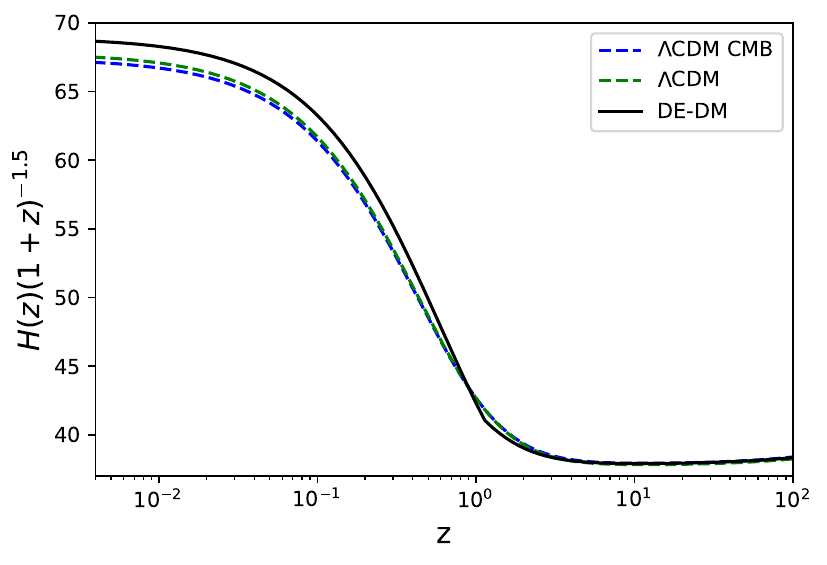}
\caption{Planck+SN+DESI+SH0ES --- Plot of $H(z)$ from Eqs.~(\ref{in20}), (\ref{parfit12}) and (\ref{parfit22}) (black). The best fit for the base $\Lambda$CDM model is depicted in orange-dashed and the blue dashed curve represents the $\Lambda$CDM model constrained by Planck. (For interpretation of the references to color in this figure legend, the reader is referred to the web version of this article.)}
\label{figHzfitP}
\end{figure}

The best-fit values including also the CMB Planck data are as follows:
\begin{align}
&H_0 = 68.92\,,\;\;
\Omega_m^0 = 0.30\,,\;\;
\Omega_r^0 =8.21\times10^{-5}\,,\label{parfit12}\\
&\Delta_{DE}^0 = 0.10\,,\quad
\bar{\Gamma} = 2.74\,,\quad
\delta = 0.79\,,\label{parfit22}
\end{align}
which give $\Delta_0 = 0.18$. For this case, the best fit for a base $\Lambda$CDM model yields: $H_0 = 67.78$, $\Omega_m^0 = 0.31$, and $\Omega_r^0 = 8.38 \times 10^{-5}$. The evolution of the Hubble parameter defined by our model is depicted in Fig.~\ref{figHzfitP}, where we also outline the $\Lambda$CDM counterpart. In the figure, we now overplot, as a reference, the profile of the $\Lambda$CDM model constrained by Planck when addressing only the CMB dataset, i.e., $H_{0\textrm{P}}$ and $\Omega_{0\textrm{P}}^m$ as previously defined. 

In this case, we see that the Hubble tension is only attenuated and the model is no longer as well-performing as in the local Universe. However, our $H(z)$ curve almost overlaps, for $z \sim 3$, with the Planck $\Lambda$CDM behavior while providing a higher value of $H_0$. For this case, we can also perform a consistency check by evaluating the expressions of $D_M^*$ using our model and the base $\Lambda$CDM profile. In this sense, we are assuming that the $\Lambda$CDM model is favored in the early Universe history, and we thus require that $D_M^*$ calculated with our $H(z)$ falls within the range $D_{M\textrm{$\Lambda$CDM}}^* \pm \Delta D_{M\textrm{$\Lambda$CDM}}^*$, with obvious notation, thereby not altering the early Universe physics with respect to a base $\Lambda$CDM model. Using the standard error propagation variance formula and the values in Tab.~\ref{table_params}, we obtain $D_M^* = 13829\,\text{Mpc}$ and $D_{M\textrm{$\Lambda$CDM}}^* = 13882 \pm 169\,\text{Mpc}$.

Concluding, we see that, including the BAO-like point at $z \simeq 1100$, the capability of the model to solve the Hubble tension is reduced, and we have just an attenuation of the tension itself, although for an appreciable amount. This suggests that, while the proposed model well describes the necessary dynamical features in the late Universe, it calls attention for a generalization when higher redshift regions are approached.

\subsubsection{General remarks}
It is worth stressing that in developing the data analysis above, we treated, as already discussed, the CMB constraints as a BAO-like point at $z \simeq 1100$. This is a well-posed choice in view of what we outlined about the setup of a late Universe model. In fact, in the proposed scenario, no change of the acoustic horizon at the recombination age is induced (actually, for $z > z_{\text{in}}$, our Hubble parameter can be translated into that emerging from a base $\Lambda$CDM behavior by a parameter redefinition only). Thus, the late Universe modification minimally affects the angular diameter distance. Under these assumptions, there is no need to use the direct Planck likelihood in the data analysis, and our model constraining procedure is quite reliable. In this respect, it is worth noting that the number of additional free parameters, in comparison to a standard $\Lambda$CDM model, is not an \emph{a priori} request, but they naturally arise from the structure of the addressed physical model. Thus, the three additional parameters we are dealing with should not be considered as an artificial procedure to adapt the Hubble parameter to the solution of the Hubble tension, but rather as the precise output of a basic phenomenological model which, capturing the main feature of DE-DM interaction, is able to interpolate apparently incompatible measurements.

Actually, when we include the Planck data, according to the scheme discussed above, the Hubble tension is no longer solved but simply slightly attenuated. The real successful feature of the present scenario is represented by its capability to accommodate different values of $H_0$ in the related $\Lambda$CDM-like fit. This significant reconciliation of different data features is reached when the SH0ES prior is taken into account, but we regard this constraint as a model-independent output of direct distance ladder implementation.

Summarizing, the results obtained above suggest that the physical concepts underlying the proposed cosmological dynamics show some promise as a possible solution to the Hubble tension problem, albeit with some additional features to be included so that the CMB data can be fitted without a marked decrement of the $H_0$ value. We infer that the decaying process of DE constituents into DM particles should probably be characterized by a more complex modulation of its intensity. In this respect, a significant role could be played by the inverse process, associated with DE constituent recombination, which will be explored in future works.

\section{Concluding remarks}
We derived a kinetic model for a DE-DM interaction, which separates the equilibrium distribution function of the two species into their equilibrium component and small corrections, accounting for the decaying process of DE constituents into DM particles. In this way, we arrive at a set of coupled equations, which are reduced to their simplified form when the DE equation of state parameter is taken as $w = -1$. This choice has been justified in order to construct a modified $\Lambda$CDM dynamics.

The model we analyzed in detail contained six free parameters. Besides $H_0$, $\Omega_m^0$ and $\Omega_r^0$, we have $\bar{\Gamma}$ (corresponding to the ratio between the Universe age and the decaying time of DE constituents), $\Delta_{DE}^0$ (which fixes the value at $z=0$ for the DE fluctuation critical parameter), and $\delta$. We first illustrated how a set of optimized parameters can provide a qualitative solution to the Hubble tension and, particularly, how the theoretical value of $H_0$ can be reconciled with the value provided by the SH0ES collaboration while the early Universe mimics the Planck $\Lambda$CDM model dynamics.

We then conducted a detailed Bayesian parameter inference procedure for both the $\Lambda$CDM model and DE-DM using Nested Sampling and several of the most recent cosmological background likelihoods. This analysis revealed that, in all cases, our model's distinctive feature is preferred by the data when considering only the fit. The Bayes' factor showed that our model is at least as good as or slightly better than $\Lambda$CDM in explaining the data, which is surprising given the additional free parameters involved. Finally, the cases where CMB background information was utilized are of particular interest too, as our model achieves a better fit to the data and slightly reduces the Hubble tension due to its unique characteristics. We take these results as motivation to study models with similar traits, as they show significance potential to address some of the problems in modern cosmology.


\begin{thebibliography}{100}
\ProvideTextCommand{\guillemotleft}{OT1}{%
  \leavevmode\raise .27ex\hbox{$\scriptscriptstyle\ll$}}
\ProvideTextCommand{\guillemotright}{OT1}{%
  \leavevmode\raise .27ex\hbox{$\scriptscriptstyle\gg$}}
\newcommand{\enquote}[1]{\guillemotleft#1\guillemotright}

\bibitem{2000Natur.404..955D}
P.~{de Bernardis}, P.~A.~R. {Ade}, J.~J. {Bock}, J.~R. {Bond}, J.~{Borrill}, A.~{Boscaleri}, K.~{Coble}, B.~P. {Crill}, G.~{De Gasperis}, P.~C. {Farese}, P.~G. {Ferreira}, K.~{Ganga}, M.~{Giacometti}, E.~{Hivon}, V.~V. {Hristov}, A.~{Iacoangeli}, A.~H. {Jaffe}, A.~E. {Lange}, L.~{Martinis}, S.~{Masi}, P.~V. {Mason}, P.~D. {Mauskopf}, A.~{Melchiorri}, L.~{Miglio}, T.~{Montroy}, C.~B. {Netterfield}, E.~{Pascale}, F.~{Piacentini}, D.~{Pogosyan}, S.~{Prunet}, S.~{Rao}, G.~{Romeo}, J.~E. {Ruhl}, F.~{Scaramuzzi}, D.~{Sforna} and N.~{Vittorio}, \emph{Nature} \textbf{404}, 955 (2000).

\bibitem{2013ApJS..208...20B}
C.~L. {Bennett}, D.~{Larson}, J.~L. {Weiland}, N.~{Jarosik}, G.~{Hinshaw}, N.~{Odegard}, K.~M. {Smith}, R.~S. {Hill}, B.~{Gold}, M.~{Halpern}, E.~{Komatsu}, M.~R. {Nolta}, L.~{Page}, D.~N. {Spergel}, E.~{Wollack}, J.~{Dunkley}, A.~{Kogut}, M.~{Limon}, S.~S. {Meyer}, G.~S. {Tucker} and E.~L. {Wright}, \emph{ApJ S.R.} \textbf{208}, 20 (2013).

\bibitem{Planck:2018vyg}
N.~Aghanim \emph{et~al.}, \emph{A\& A} \textbf{641}, A6 (2020), [Erratum: Astron.Astrophys. 652, C4 (2021)].

\bibitem{2024ApJ...963L..43A}
Richard~I. {Anderson}, Nolan~W. {Koblischke} and Laurent {Eyer}, \emph{ApJ Lett.} \textbf{963}, L43 (2024).

\bibitem{Scolnic_2022}
Dan~M. Scolnic, Dillon Brout, Anthony Carr, Adam~G. Riess, Tamara~M. Davis, Arianna Dwomoh, David~O. Jones, Noor Ali, Pranav Charvu, Rebecca Chen, Erik~R. Peterson, Brodie Popovic, Benjamin~M. Rose, Charlotte~M. Wood, Peter~J. Brown, Ken Chambers, David~A. Coulter, Kyle~G. Dettman, Georgios Dimitriadis, Alexei~V. Filippenko, Ryan~J. Foley, Saurabh~W. Jha, Charles~D. Kilpatrick, Robert~P. Kirshner, Yen-Chen Pan, Armin Rest, Cesar Rojas-Bravo, Matthew~R. Siebert, Benjamin~E. Stahl and WeiKang Zheng, \emph{ApJ} \textbf{938}, 113 (2022).

\bibitem{2018ApJ...859..101S}
D.~M. {Scolnic}, D.~O. {Jones}, A.~{Rest}, Y.~C. {Pan}, R.~{Chornock}, R.~J. {Foley}, M.~E. {Huber}, R.~{Kessler}, G.~{Narayan}, A.~G. {Riess}, S.~{Rodney}, E.~{Berger}, D.~J. {Brout}, P.~J. {Challis}, M.~{Drout}, D.~{Finkbeiner}, R.~{Lunnan}, R.~P. {Kirshner}, N.~E. {Sanders}, E.~{Schlafly}, S.~{Smartt}, C.~W. {Stubbs}, J.~{Tonry}, W.~M. {Wood-Vasey}, M.~{Foley}, J.~{Hand}, E.~{Johnson}, W.~S. {Burgett}, K.~C. {Chambers}, P.~W. {Draper}, K.~W. {Hodapp}, N.~{Kaiser}, R.~P. {Kudritzki}, E.~A. {Magnier}, N.~{Metcalfe}, F.~{Bresolin}, E.~{Gall}, R.~{Kotak}, M.~{McCrum} and K.~W. {Smith}, \emph{ApJ} \textbf{859}, 101 (2018).

\bibitem{Brout:2022vxf}
Dillon Brout \emph{et~al.}, \emph{ApJ} \textbf{938}, 110 (2022).

\bibitem{DAINOTTI202430}
M.G. Dainotti, G.~Bargiacchi, M.~Bogdan, S.~Capozziello and S.~Nagataki, \emph{Journal of High Energy Astrophysics} \textbf{41}, 30 (2024).

\bibitem{Dainotti2023ApJ...951...63D}
Maria~Giovanna {Dainotti}, Giada {Bargiacchi}, Malgorzata {Bogdan}, Aleksander~Lukasz {Lenart}, Kazunari {Iwasaki}, Salvatore {Capozziello}, Bing {Zhang} and Nissim {Fraija}, \emph{ApJ} \textbf{951}, 63 (2023).

\bibitem{Riess:2021jrx}
Adam~G. Riess \emph{et~al.}, \emph{Astrophys. J. Lett.} \textbf{934}, L7 (2022).

\bibitem{Scolnic:2023mrv}
D.~Scolnic, A.~G. Riess, J.~Wu, S.~Li, G.~S. Anand, R.~Beaton, S.~Casertano, R.~I. Anderson, S.~Dhawan and X.~Ke, \emph{Astrophys. J. Lett.} \textbf{954}, L31 (2023).

\bibitem{Jones:2022mvo}
D.~O. Jones \emph{et~al.}, \emph{Astrophys. J.} \textbf{933}, 172 (2022).

\bibitem{Anand:2021sum}
Gagandeep~S. Anand, R.~Brent Tully, Luca Rizzi, Adam~G. Riess and Wenlong Yuan, \emph{Astrophys. J.} \textbf{932}, 15 (2022).

\bibitem{Freedman:2021ahq}
Wendy~L. Freedman, \emph{Astrophys. J.} \textbf{919}, 16 (2021).

\bibitem{Uddin:2023iob}
Syed~A. Uddin \emph{et~al.}, \emph{arXiv:2308.01875}  (2023).

\bibitem{Huang:2023frr}
Caroline~D. Huang \emph{et~al.}, \emph{Astrophys. J.} \textbf{963}, 83 (2024).

\bibitem{Li:2024yoe}
Siyang Li, Adam~G. Riess, Stefano Casertano, Gagandeep~S. Anand, Daniel~M. Scolnic, Wenlong Yuan, Louise Breuval and Caroline~D. Huang, \emph{Astrophys. J.} \textbf{966}, 20 (2024).

\bibitem{Pesce:2020xfe}
D.~W. Pesce \emph{et~al.}, \emph{Astrophys. J. Lett.} \textbf{891}, L1 (2020).

\bibitem{Kourkchi:2020iyz}
Ehsan Kourkchi, R.~Brent Tully, Gagandeep~S. Anand, Helene~M. Courtois, Alexandra Dupuy, James~D. Neill, Luca Rizzi and Mark Seibert, \emph{Astrophys. J.} \textbf{896}, 3 (2020).

\bibitem{Schombert:2020pxm}
James Schombert, Stacy McGaugh and Federico Lelli, \emph{Astron. J.} \textbf{160}, 71 (2020).

\bibitem{Blakeslee:2021rqi}
John~P. Blakeslee, Joseph~B. Jensen, Chung-Pei Ma, Peter~A. Milne and Jenny~E. Greene, \emph{Astrophys. J.} \textbf{911}, 65 (2021).

\bibitem{deJaeger:2022lit}
T.~de~Jaeger, L.~Galbany, A.~G. Riess, B.~E. Stahl, B.~J. Shappee, A.~V. Filippenko and W.~Zheng, \emph{Mon. Not. Roy. Astron. Soc.} \textbf{514}, 4620 (2022).

\bibitem{Murakami:2023xuy}
Yukei~S. Murakami, Adam~G. Riess, Benjamin~E. Stahl, W.~D'Arcy Kenworthy, Dahne-More~A. Pluck, Antonella Macoretta, Dillon Brout, David~O. Jones, Dan~M. Scolnic and Alexei~V. Filippenko, \emph{JCAP} \textbf{11}, 046 (2023).

\bibitem{Breuval:2024lsv}
Louise Breuval, Adam~G. Riess, Stefano Casertano, Wenlong Yuan, Lucas~M. Macri, Martino Romaniello, Yukei~S. Murakami, Daniel Scolnic, Gagandeep~S. Anand and Igor Soszy\'nski, \emph{arXiv:2404.08038}  (2024).

\bibitem{SPT-3G:2022hvq}
L.~Balkenhol \emph{et~al.}, \emph{Phys. Rev. D} \textbf{108}, 023510 (2023).

\bibitem{ACT:2020gnv}
Simone Aiola \emph{et~al.}, \emph{JCAP} \textbf{12}, 047 (2020).

\bibitem{2024A&A...686A.210F}
Horst {Foidl} and Tanja {Rindler-Daller}, \emph{A\& A} \textbf{686}, A210 (2024).

\bibitem{2024arXiv241218493L}
Orlando {Luongo} and Marco {Muccino}, \emph{arXiv e-prints} arXiv:2412.18493 (2024).

\bibitem{2024arXiv241214750S}
Rahul {Shah}, Purba {Mukherjee}, Soumadeep {Saha}, Utpal {Garain} and Supratik {Pal}, \emph{arXiv e-prints} arXiv:2412.14750 (2024).

\bibitem{2021ApJ...914L..40D}
M.~G. {Dainotti}, V.~{Petrosian} and L.~{Bowden}, \emph{ApJ Lett.} \textbf{914}, L40 (2021).

\bibitem{Dainotti2023mnras}
M.~G. {Dainotti}, A.~{\L}. {Lenart}, A.~{Chraya}, G.~{Sarracino}, S.~{Nagataki}, N.~{Fraija}, S.~{Capozziello} and M.~{Bogdan}, \emph{Mon. Not. RAS} \textbf{518}, 2201 (2023).

\bibitem{2022PASJ...74.1095D}
Maria~Giovanna {Dainotti}, Giuseppe {Sarracino} and Salvatore {Capozziello}, \emph{Pub. Astro. Soc. Japan} \textbf{74}, 1095 (2022).

\bibitem{2024PDU....101874}
Giovanni {Montani}, Nakia {Carlevaro} and Maria~Giovanna {Dainotti}, \emph{Phys. Dark Univ.} \textbf{48}, 101847 (2024).

\bibitem{DiValentino:2021izs}
Eleonora Di~Valentino, Olga Mena, Supriya Pan, Luca Visinelli, Weiqiang Yang, Alessandro Melchiorri, David~F. Mota, Adam~G. Riess and Joseph Silk, \emph{Class. Quant. Grav.} \textbf{38}, 153001 (2021).

\bibitem{2024Univ...10..140C}
Salvatore {Capozziello}, Giuseppe {Sarracino} and Giulia {De Somma}, \emph{Universe} \textbf{10}, 140 (2024).

\bibitem{2023PDU....4001201C}
Salvatore {Capozziello}, Giuseppe {Sarracino} and Alessandro D.~A.~M. {Spallicci}, \emph{Phys. Dark Univ.} \textbf{40}, 101201 (2023).

\bibitem{Kamionkowski:2022pkx}
Marc Kamionkowski and Adam~G. Riess, \emph{Ann. Rev. Nucl. Part. Sci.} \textbf{73}, 153 (2023).

\bibitem{Khalife:2023qbu}
Ali~Rida Khalife, Maryam~Bahrami Zanjani, Silvia Galli, Sven G\"unther, Julien Lesgourgues and Karim Benabed, \emph{arXiv:2312.09814}  (2023).

\bibitem{Abdalla:2022yfr}
Elcio Abdalla \emph{et~al.}, \emph{JHEAp} \textbf{34}, 49 (2022).

\bibitem{Perivolaropoulos:2021jda}
Leandros Perivolaropoulos and Foteini Skara, \emph{New Astron. Rev.} \textbf{95}, 101659 (2022).

\bibitem{DiValentino:2020zio}
Eleonora Di~Valentino \emph{et~al.}, \emph{Astropart. Phys.} \textbf{131}, 102605 (2021).

\bibitem{Verde:2023lmm}
Licia Verde, Nils Sch\"oneberg and H\'ector Gil-Mar\'\i{}n, \emph{arXiv e-prints}  (2023).

\bibitem{Giare:2024syw}
William Giar\`e, Jonathan Betts, Carsten van~de Bruck and Eleonora Di~Valentino, \emph{arXiv:2406.07493}  (2024).

\bibitem{Giare:2024akf}
William Giar\`e, \emph{Phys. Rev. D} \textbf{109}, 123545 (2024).

\bibitem{2024arXiv240814878B}
S.~{Banerjee} and A.~{Paul}, \emph{arXiv e-prints} arXiv:2408.14878 (2024).

\bibitem{hu-wang03}
X.D. {Jia}, J.P. {Hu} and F.Y. {Wang}, \emph{Universe} \textbf{9}, 94 (2023).

\bibitem{arXiv:2411.16678}
S.H. {Mirpoorian}, K.~{Jedamzik} and L~{Pogosian}, \emph{arXiv:2411.16678}  (2024).

\bibitem{arXiv:2304.01831}
R.S. {Miller}, \emph{arXiv:2304.01831}  (2024).

\bibitem{Dainotti2021apj-powerlaw}
Maria~Giovanna Dainotti, Biagio De~Simone, Tiziano Schiavone, Giovanni Montani, Enrico Rinaldi and Gaetano Lambiase, \emph{ApJ} \textbf{912}, 150 (2021).

\bibitem{Dainottigalaxies10010024}
Maria~Giovanna Dainotti, Biagio De~Simone, Tiziano Schiavone, Giovanni Montani, Enrico Rinaldi, Gaetano Lambiase, Malgorzata Bogdan and Sahil Ugale, \emph{Galaxies} \textbf{10} (2022).

\bibitem{Krishnan:2020obg}
C.~Krishnan, Eoin~\'O. Colg\'ain, Ruchika, Anjan~A. Sen, M.~M. Sheikh-Jabbari and Tao Yang, \emph{Phys. Rev. D} \textbf{102}, 103525 (2020).

\bibitem{Krishnan:2020vaf}
C.~Krishnan, E.~\'O. Colg\'ain, M.~M. Sheikh-Jabbari and Tao Yang, \emph{Phys. Rev. D} \textbf{103}, 103509 (2021).

\bibitem{kazantzidis}
L.~Kazantzidis and L.~Perivolaropoulos, \emph{Phys. Rev. D} \textbf{102}, 023520 (2020).

\bibitem{hu-wang02}
X.D. {Jia}, J.P. {Hu} and F.Y. {Wang}, \emph{A \& A} \textbf{674}, A45 (2023).

\bibitem{hu-wang01}
J.P. {Hu} and F.Y. {Wang}, \emph{Mon. Not. RAS} \textbf{517}, 576 (2022).

\bibitem{2025arXiv250111772D}
Maria~Giovanna {Dainotti}, Biagio {De Simone}, Anargha {Mondal}, Kazunori {Kohri}, Augusto {C{\'e}sar Caligula do Esp{\'\i}rito Santo Pedreira}, Angel {Bashyal}, Rafid~Hasan {Dejrah}, Shigehiro {Nagataki}, Giovanni {Montani}, Anurag {Singh}, Ayush {Garg}, Mudit {Parakh}, Nissim {Fraija}, Rohit {Mandal}, Hrikhes {Sarkar}, Tasneem {Jareen}, Kinshuk {Jarial} and Gaetano {Lambiase}, \emph{arXiv e-prints} arXiv:2501.11772 (2025).

\bibitem{Brout:2020bbg}
Dillon Brout, Samuel Hinton and Daniel Scolnic, \emph{Astrophys. J. Lett.} \textbf{912}, L26 (2021).

\bibitem{Sotiriou-Faraoni:2010}
Thomas~P. Sotiriou and Valerio Faraoni, \emph{Rev. Mod. Phys.} \textbf{82}, 451 (2010).

\bibitem{NOJIRI201159}
Shin’ichi Nojiri and Sergei~D. Odintsov, \emph{Phys. Rept.} \textbf{505}, 59 (2011).

\bibitem{arx1705.11098}
S.~{Nojiri}, S.~D. {Odintsov} and V.~K. {Oikonomou}, \emph{Phys. Rept.} \textbf{692}, 1 (2017).

\bibitem{arx2307.16308}
Sergei~D. {Odintsov}, Vasilis~K. {Oikonomou}, Ifigeneia {Giannakoudi}, Fotis~P. {Fronimos} and Eirini~C. {Lymperiadou}, \emph{Symmetry} \textbf{15}, 1701 (2023).

\bibitem{2011PhR...509..167C}
Salvatore {Capozziello} and Mariafelicia {de Laurentis}, \emph{Phys. Rept.} \textbf{509}, 167 (2011).

\bibitem{schiavone_mnras}
Tiziano Schiavone, Giovanni Montani and Flavio Bombacigno, \emph{Mon. Not. RAS} \textbf{522}, L72 (2023).

\bibitem{Nojiri:2022ski}
S.~Nojiri, S.~D. Odintsov and V.~K. Oikonomou, \emph{Nucl. Phys. B} \textbf{980}, 115850 (2022).

\bibitem{Odintsov:2020qzd}
Sergei~D. Odintsov, Diego S\'aez-Chill\'on~G\'omez and German~S. Sharov, \emph{Nucl. Phys. B} \textbf{966}, 115377 (2021).

\bibitem{Petronikolou:2023cwu}
Maria Petronikolou and Emmanuel~N. Saridakis, \emph{Universe} \textbf{9}, 397 (2023).

\bibitem{FrancoAbellan:2023gec}
Guillermo Franco~Abell\'an, Matteo Braglia, Mario Ballardini, Fabio Finelli and Vivian Poulin, \emph{JCAP} \textbf{12}, 017 (2023).

\bibitem{Ravi:2023nsn}
Kumar Ravi, Anirban Chatterjee, Biswajit Jana and Abhijit Bandyopadhyay, \emph{Mon. Not. Roy. Astron. Soc.} \textbf{527}, 7626 (2024).

\bibitem{deangelis-fr-mnras}
Giovanni {Montani}, Mariaveronica {De Angelis}, Flavio {Bombacigno} and Nakia {Carlevaro}, \emph{Mon. Not. RAS} \textbf{527}, L156–L161 (2024).

\bibitem{2024PhRvD.109b3527A}
S.A. {Adil}, {\"O}.~{Akarsu}, E.~{Di Valentino}, R.C. {Nunes}, E.~{{\"O}z{\"u}lker}, A.A. {Sen} and E.~{Specogna}, \emph{Phys. Rev. D} \textbf{109}, 023527 (2024).

\bibitem{2024PDU....4401486M}
Giovanni {Montani}, Nakia {Carlevaro} and Maria~Giovanna {Dainotti}, \emph{Phys. Dark Univ.} \textbf{44}, 101486 (2024).

\bibitem{erdem24a}
Recai {Erdem}, \emph{arXiv e-prints} arXiv:2402.16791 (2024).

\bibitem{eBOSS:2020yzd}
Shadab Alam \emph{et~al.}, \emph{Phys. Rev. D} \textbf{103}, 083533 (2021).

\bibitem{DESI:2024mwx}
A.~G. Adame \emph{et~al.}, \emph{arXiv:2404.03002}  (2023).

\bibitem{2021MNRAS.505.3866E}
George {Efstathiou}, \emph{Mon. Not. RAS} \textbf{505}, 3866 (2021).

\bibitem{Knox:2019rjx}
Lloyd Knox and Marius Millea, \emph{Phys. Rev. D} \textbf{101}, 043533 (2020).

\bibitem{2022JHEAp..36...27V}
Sunny {Vagnozzi}, Fabio {Pacucci} and Abraham {Loeb}, \emph{Journal of High Energy Astrophysics} \textbf{36}, 27 (2022).

\bibitem{PhysRevD.98.083501}
Sunny Vagnozzi, Suhail Dhawan, Martina Gerbino, Katherine Freese, Ariel Goobar and Olga Mena, \emph{Phys. Rev. D} \textbf{98}, 083501 (2018).

\bibitem{2023Univ....9..393V}
Sunny {Vagnozzi}, \emph{Universe} \textbf{9}, 393 (2023).

\bibitem{2020PhRvD.102b3518V}
Sunny {Vagnozzi}, \emph{Phys. Rev. D} \textbf{102}, 023518 (2020).

\bibitem{Allali:2021azp}
Itamar~J. Allali, Mark~P. Hertzberg and Fabrizio Rompineve, \emph{Phys. Rev. D} \textbf{104}, L081303 (2021).

\bibitem{Anchordoqui:2021gji}
Luis~A. Anchordoqui, Eleonora Di~Valentino, Supriya Pan and Weiqiang Yang, \emph{JHEAp} \textbf{32}, 28 (2021).

\bibitem{Khosravi:2021csn}
Nima Khosravi and Marzieh Farhang, \emph{Phys. Rev. D} \textbf{105}, 063505 (2022).

\bibitem{Clark:2021hlo}
Steven~J. Clark, Kyriakos Vattis, JiJi Fan and Savvas~M. Koushiappas, \emph{Phys. Rev. D} \textbf{107}, 083527 (2023).

\bibitem{Wang:2022jpo}
Hao Wang and Yun-Song Piao, \emph{Phys. Lett. B} \textbf{832}, 137244 (2022).

\bibitem{Anchordoqui:2022gmw}
Luis~A. Anchordoqui, Vernon Barger, Danny Marfatia and Jorge~F. Soriano, \emph{Phys. Rev. D} \textbf{105}, 103512 (2022).

\bibitem{Reeves:2022aoi}
Alexander Reeves, Laura Herold, Sunny Vagnozzi, Blake~D. Sherwin and Elisa G.~M. Ferreira, \emph{Mon. Not. Roy. Astron. Soc.} \textbf{520}, 3688 (2023).

\bibitem{Yao:2023qve}
Yan-Hong Yao and Xin-He Meng, \emph{Commun. Theor. Phys.} \textbf{76}, 075401 (2024).

\bibitem{daCosta:2023mow}
Simony~Santos da~Costa, D\^eivid~R. da~Silva, \'Alvaro~S. de~Jesus, Nelson Pinto-Neto and Farinaldo~S. Queiroz, \emph{JCAP} \textbf{04}, 035 (2024).

\bibitem{Wang:2024dka}
Hao Wang and Yun-Song Piao, \enquote{{Dark energy in light of recent DESI BAO and Hubble tension}},  (2024).

\bibitem{Ziad23}
Z.~{Sakr}, \emph{Phys. Rev. D} \textbf{108}, 083519 (2023).

\bibitem{Dainotti2024PDU....4401428D}
M.~G. {Dainotti}, A.~{\L}. {Lenart}, M.~Ghodsi {Yengejeh}, S.~{Chakraborty}, N.~{Fraija}, E.~{Di Valentino} and G.~{Montani}, \emph{Physics of the Dark Universe} \textbf{44}, 101428 (2024).

\bibitem{arx2203.10558}
E.~{{\'O} Colg{\'a}in}, M.~M. {Sheikh-Jabbari}, R.~{Solomon}, G.~{Bargiacchi}, S.~{Capozziello}, M.~G. {Dainotti} and D.~{Stojkovic}, \emph{Phys. Rev. D.} \textbf{106}, L041301 (2022).

\bibitem{arx2206.11447}
E.~{{\'O} Colg{\'a}in}, M.~M. {Sheikh-Jabbari}, R.~{Solomon}, M.~G. {Dainotti} and D.~{Stojkovic}, \emph{Physics of the Dark Universe} \textbf{44}, 101464 (2024).

\bibitem{arx2304.02718}
Ruair{\'\i} {Mc Conville} and Eoin {{\'O} Colg{\'a}in}, \emph{Phys. Rev. D} \textbf{108}, 123533 (2023).

\bibitem{2016RPPh...79i6901W}
B.~{Wang}, E.~{Abdalla}, F.~{Atrio-Barandela} and D.~{Pav{\'o}n}, \emph{Rep. Prog. Phys.} \textbf{79}, 096901 (2016).

\bibitem{naidoo2024PhRvD}
Krishna {Naidoo}, Mariana {Jaber}, Wojciech~A. {Hellwing} and Maciej {Bilicki}, \emph{Phys. Rev. D} \textbf{109}, 083511 (2024).

\bibitem{Pourtsidou:2016ico}
Alkistis Pourtsidou and Thomas Tram, \emph{Phys. Rev. D} \textbf{94}, 043518 (2016).

\bibitem{DiValentino:2017iww}
Eleonora Di~Valentino, Alessandro Melchiorri and Olga Mena, \emph{Phys. Rev. D} \textbf{96}, 043503 (2017).

\bibitem{Kumar:2017dnp}
Suresh Kumar and Rafael~C. Nunes, \emph{Phys. Rev. D} \textbf{96}, 103511 (2017).

\bibitem{Yang:2018uae}
Weiqiang Yang, Ankan Mukherjee, Eleonora Di~Valentino and Supriya Pan, \emph{Phys. Rev. D} \textbf{98}, 123527 (2018).

\bibitem{vonMarttens:2019ixw}
Rodrigo von Marttens, Lucas Lombriser, Martin Kunz, Valerio Marra, Luciano Casarini and Jailson Alcaniz, \emph{Phys. Dark Univ.} \textbf{28}, 100490 (2020).

\bibitem{Lucca:2020zjb}
Matteo Lucca and Deanna~C. Hooper, \emph{Phys. Rev. D} \textbf{102}, 123502 (2020).

\bibitem{Zhai:2023yny}
Yuejia Zhai, William Giar\`e, Carsten van~de Bruck, Eleonora Di~Valentino, Olga Mena and Rafael~C. Nunes, \emph{JCAP} \textbf{07}, 032 (2023).

\bibitem{Bernui:2023byc}
Armando Bernui, Eleonora Di~Valentino, William Giar\`e, Suresh Kumar and Rafael~C. Nunes, \emph{Phys. Rev. D} \textbf{107}, 103531 (2023).

\bibitem{Hoerning:2023hks}
Gabriel~A. Hoerning, Ricardo~G. Landim, Luiza~O. Ponte, Raphael~P. Rolim, Filipe~B. Abdalla and Elcio Abdalla, \emph{arXiv:2308.05807}  (2023).

\bibitem{Giare:2024ytc}
William Giar\`e, Yuejia Zhai, Supriya Pan, Eleonora Di~Valentino, Rafael~C. Nunes and Carsten van~de Bruck, \emph{arXiv:2404.02110}  (2024).

\bibitem{Giare:2024smz}
William Giar\`e, Miguel~A. Sabogal, Rafael~C. Nunes and Eleonora Di~Valentino, \emph{arXiv:2404.15232}  (2024).

\bibitem{escamilla2023idegp}
Luis~A Escamilla, {\"O}zg{\"u}r Akarsu, Eleonora Di~Valentino and J~Alberto Vazquez, \emph{Journal of Cosmology and Astroparticle Physics} \textbf{2023}, 051 (2023).

\bibitem{Benisty:2024lmj}
David Benisty, Supriya Pan, Denitsa Staicova, Eleonora Di~Valentino and Rafael~C. Nunes, \emph{arXiv:2403.00056}  (2024).

\bibitem{Silva:2024ift}
Emanuelly Silva, Ubaldo Z\'u\~niga Bola\~no, Rafael~C. Nunes and Eleonora Di~Valentino, \emph{arXiv:2403.19590}  (2024).

\bibitem{Forconi:2023hsj}
Matteo Forconi, William Giar\`e, Olga Mena, Ruchika, Eleonora Di~Valentino, Alessandro Melchiorri and Rafael~C. Nunes, \emph{JCAP} \textbf{05}, 097 (2024).

\bibitem{DiValentino:2019ffd}
Eleonora Di~Valentino, Alessandro Melchiorri, Olga Mena and Sunny Vagnozzi, \emph{Phys. Dark Univ.} \textbf{30}, 100666 (2020).

\bibitem{arXiv:2409.15513}
A.~S. de~Jesus, M.M.A. Paixao, D.R. da~Silva, F.S. Queiroz and N.~Pinto-Neto, \emph{arXiv:2409.15513}  (2024).

\bibitem{spin12}
F.~Hammad, R.~Saadati, M.~Simard and S.~Novoa-Cattivelli, \emph{arXiv:2410.01787}  (2024).

\bibitem{bib:montani-primordialcosmology}
G.~Montani, M.~V. Battisti, R.~Benini and G.~Imponente, \emph{Primordial Cosmology} (World Scientific, Singapore) (2011).

\bibitem{1988PhRvL..61.1446M}
Michael~S. {Morris}, Kip~S. {Thorne} and Ulvi {Yurtsever}, \emph{Phys. Rev. Lett.} \textbf{61}, 1446 (1988).

\bibitem{melchiorriDE}
Alessandro {Melchiorri}, Luca {Pagano} and Stefania {Pandolfi}, \emph{Phys. Rev. D} \textbf{76}, 041301 (2007).

\bibitem{2008ARA&A..46..385F}
J.~A. {Frieman}, M.~S. {Turner} and D.~{Huterer}, \emph{Annual Rev. Astron. Astrophys.} \textbf{46}, 385 (2008).

\bibitem{simplemc}
{A. Slosar and J. A. Vazquez}, \url{https://github.com/ja-vazquez/SimpleMC}.

\bibitem{speagle2020dynesty}
Joshua~S. Speagle, \emph{Mon. Not. Roy. Astron. Soc.} \textbf{493}, 3132 (2020).

\bibitem{Trotta:2008qt}
Roberto Trotta, \emph{Contemp. Phys.} \textbf{49}, 71 (2008).

\bibitem{Cooke:2013cba}
Ryan Cooke, Max Pettini, Regina~A. Jorgenson, Michael~T. Murphy and Charles~C. Steidel, \emph{Astrophys. J.} \textbf{781}, 31 (2014).

\bibitem{BOSS:2014hhw}
\'Eric Aubourg \emph{et~al.}, \emph{Phys. Rev. D} \textbf{92}, 123516 (2015).

\bibitem{Kamionkowski23}
Marc {Kamionkowski} and Adam~G. {Riess}, \emph{Annual Review of Nuclear and Particle Science} \textbf{73}, 153 (2023).

\end{thebibliography}

\end{document}